\documentclass[preprintnumbers,amsmath,amssymb]{revtex4}
\usepackage{graphicx}
\begin{document}
\preprint{IMAFF-RCA-05-02}
\title{Holographic cosmic energy, fundamental theories and the future of the universe}
\author{Pedro F. Gonz\'{a}lez-D\'{\i}az}
\affiliation{Colina de los Chopos, Centro de F\'{\i}sica ``Miguel A.
Catal\'{a}n'', Instituto de Matem\'{a}ticas y F\'{\i}sica Fundamental,\\
Consejo Superior de Investigaciones Cient\'{\i}ficas, Serrano 121,
28006 Madrid (SPAIN)}
\date{\today}
\begin{abstract}
Holographic dark energy models have been recently suggested which
most clearly show that accelerating cosmology appears to be
incompatible with mathematically consistent formulations of
fundamental theories such as string/M theories. In this paper it is
however argued that holographic phantom models are no longer
incompatible with such theories, provided that we allow for the
existence of wormholes and ringholes near the big rip singularity
and a quantization condition on the parameter of the equation of
state is introduced. It is also seen that such a condition actually
implies a quantization of the phantom energy which stabilizes the
fluid against decay processes.

\end{abstract}

\maketitle


\vspace{1cm}


\pagebreak

\section{Introduction}

The last years have been really exciting in cosmology. The
discovery that the universe is currently accelerating [1] has
opened a true plethora de developments which have much attracted
researchers from other fields, including string theorists and
astrophysicists. Among these developments one cannot forget the
implication that Einstein gravity cannot by itself govern the
evolution of the universe, which has driven two distinct research
avenues. On the one hand, some cosmologists have searched modified
(generalized) gravitational theories in which the Hilbert-Einstein
action is supplemented with inverse curvature terms able to
predict the late time acceleration [2], and on the other hand,
others have looked at a scenario where Einstein gravity is
preserved but some vacuum fields are introduced to account for the
supernova Ia observations [3]. The energy associated with these
fields has been dubbed dark energy and, if it exists it would do
so in such a way that it becomes currently dominating over all
other kinds of energy in the universe. Of particular interest are
the so-called quintessence fields [4]. These are equivalent to a
cosmic fluid characterized by an equation of state, $p=w\rho$,
where the energy density $\rho$ is positive and the index $w$ has
been constrained by observations to be within the interval
$-0.8\leq w\leq -1.4$ [5]. The resulting negative pressure would
then play the role of an anti-gravity regime which is ultimately
responsible for the late-time acceleration of the universe. One of
the greatest challenges posed by this discovery is its apparent
essential incompatibility with those fundamental theories which
rely on the existence of a S-matrix or S-vector formulation, that
is quantum-gravity and string/M theories [6]. However, more recent
work has made great strides toward understanding how de Sitter
solutions may be eventually accommodated in a string theory
framework, so largely weakening the essential character of the
above puzzle, at least in the particular cases being considered.
Furthermore, it looks straightforward enough to envisage a
universe which accelerates only temporarily, until the dark energy
decays e.g via some phase transition, after which it approached
Minkowski space where the above puzzle would vanish. Nevertheless,
it is also straightforward enough to imagine that the universe
will again accelerate on some period of the future where the
puzzle would once again reappears. Actually, Caldwell et al [7]
have now proposed a model which quite comfortably accommodate
current observations where the future phase transition precisely
leads to a phantom universe phase. Thus, even though it perhaps
hardly could be maintained as being a strictly essential problem,
the solution to the above cosmic-fundamental theory puzzle appears
still be a necessary requirement.

This cosmic-fundamental theory puzzle -where the word
"fundamental" stands here for emphasizing the basic character of
the involved field theories- is most apparent when dark energy
models are combined with holographic bounds on entropy [8] to
yield up a cosmic theory where the short distance UV cutoff and
the large distance IR cutoff are mutually related [9]. A
holographic model of dark energy has been recently suggested by Li
[10] which is based on the following relation between the Hubble
parameter and a suitably chosen holographic screen,
\begin{equation}
H^2=\frac{\dot{a}^2}{a^2}=\frac{8\pi G\rho}{3}=\frac{c^2}{R_h^2},
\end{equation}
where $c$ is a numerical parameter which is related to the index $w$
by $w=-(1+2/c)/3$, $a\equiv a(t)$ is the scale factor and
$R_h=a\int_t^{\infty}dt/a(t)$ is the proper size of the future event
horizon which is introduced to play the role of the cosmic
holographic surface. This integral expression corresponds to the
known definition of an event horizon, but its use as a holographic
screen makes that definition equivalent through Eq. (1) to the
Hubble radius $H^{-1}$, in spite of the feature that the actual
cosmic horizon can be much larger if an early inflationary period is
considered.

Thus, for a scale factor given by
$a=\left(a_0^{1-1/c}+(1-1/c)\sqrt{\frac{8\pi
G\rho_0}{3}}(t-t_0)\right)^{1/(1-1/c)}=T(t)^{1/(1-1/c)}$, with $a_0$
and $\rho_0$ the scale factor and the energy density, respectively,
at the initial time $t_0$, in a generic quintessence model, the
proper size of the future event horizon is given by [10]
\begin{equation}
R_h=- cT(t)^{1/(1-1/c)}\left(\left.
T(t')^{-1/[c(1-1/c)]}\right|_t^{\infty} \right) .
\end{equation}
Clearly, if $c>1\rightarrow w>-1$ then $R_h=cT(t)$, which is finite
for finite $t$. This is the Li proposal [10] for dark energy which
becomes well defined only when $w>-1$. For such a regime, all the
information contained in the universe at time $t$ would thereby be
encoded at the finite surface on the future event horizon. Now,
since no information whatsoever on events taking place beyond that
horizon could be obtained by any observer at $t$, theories based on
the S-matrix that relate points infinitely separated spatially could
in principle not be consistently formulated within this context.

In this paper we shall discuss in more detail models of
holographic phantom cosmology in relation with the possibility of
formulating mathematically consistent fundamental theories,
extending the above analysis to other dark energy models. It will
also be argued that in order for this possibility to be maintained
it is necessary to quantize the index of the equation of state,
$p=w\rho$, and hence the energy of the phantom universe. The
resulting quantization of phantom energy may become a key
ingredient for phantom energy stabilization, as it forbids any
conversion process involving the emission of arbitrarily small
amounts of energy which do not satisfy the quantization
requirement.

\section{A holographic phantom energy model}

A solution to the cosmic-fundamental theory problem could be
nevertheless derived if we allow $w$ to take on values less than
-1 [11]. Actually, for $w<-1$ (i.e. for $c<1$) it follows from Eq.
(2) that the proper size of the future horizon inexorably becomes
infinity; or what is equivalent, the future event horizon will
vanish for phantom energy, so solving the above puzzle in that
case. However, a holographic phantom model cannot use Eq. (1) for
the Hubble parameter $H$ because in this case Eq. (1) leads to
$H=0$ along the whole cosmic evolution dominated by $w<-1$. Thus,
the Li proposal for holographic dark energy does not work when
$w<-1$. For in that case, instead of Eq. (1) I suggest using the
following cosmic holographic definition (see also Refs. [12] and
[13])
\begin{equation}
H^2_{ph}=\frac{\dot{a}^2}{a^2}=\frac{8\pi G\rho}{3}=
\frac{c^2}{R_*^2} ,
\end{equation}
where $R_*=a(t)\int_t^{t_*} dt'/a(t') =cT(t)$, with $t_*$ the time
at which the big rip takes place [12],
$t_*=t_0+2a_0^{-3(|w|-1)/2}/[3(|w|-1)]$. Now, $H\neq 0$ satisfying
the first equality in the previous Eq. (3). So, in the case of the
holographic phantom energy we must take for the size of the
ultimate region whose total energy is chosen not to exceed the
mass of a black hole with the same size, $L$, the proper size of
the horizon at the big rip $R_*$, i.e. $L=R_*$, instead of the
proper size of the future event horizon $R_h$, which was chosen
for holographic dark energy with $c\geq 1$. Since the cosmic
fundamental theory puzzle refers to the future event horizon, we
can then have both a well-defined holographic phantom theory and a
suitable solution to that puzzle. There is however a question
remaining. In the quintessential phantom regime that corresponds
to $w<-1$ there will be a real big rip singularity at the finite
time $t=t_*$ in the future [14]. If we want then to preserve an
infinite proper size for the event horizon, it will be later seen
that it is necessary to allow [11] for wormhole-mediated
connections between the regions before and after the big rip
singularity. These traversable tunnels could then be supporting an
unbounded flow of information from any point on the region after
the big rip to the given observer. Wormholes of different types
should actually be expected to occur in general relativity. They
in fact correspond to well-defined solutions to the Einstein
equations and, therefore, can be considered in our analysis,
mainly in models of phantom energy which is actually is the stuff
that is required to cast a suitable energy-momentum tensor giving
rise to wormhole solution [15,16]. Such wormholes are therefore
necessary ingredients in the context of the present paper.

The above usual definition of the event horizon involves the
integration limit at $t=\infty$ and so the entire region defined
for times larger than that for the rig rip is included in such a
definition for phantom energy. Although the big rip corresponds to
a curvature singularity and therefore at first sight it would mark
the end of the evolution of the universe, there are however at
least two reasons why one should also consider the region after
the big rip, $t>t_*$ to be physical. On the one hand, it has been
recently shown [16] that wormholes can be made of phantom energy
and therefore they should quite naturally crop up in the universe
provided that $w<-1$. On the other hand and more importantly, it
will be later seen that as one approaches the finite future
singularity the cosmic spacetime metric becomes expressible as the
metric on a five-dimensional hyperboloid which is invariant under
the Misner symmetry and therefore describes bounded space-time
regions where closed timelike curves can occur through traversable
wormholes and ringholes in the neighborhood of the singularity. In
both cases, such space-time tunnelings can join the regions before
and after the big rip so that information signaling may travel
from one region to another without reaching the singularity at
finite time which is shortcut in this way. The physical region for
all these signalings and the physical objects able to traverse the
traversable wormholes and ringholes will then extend beyond the
big rip singularity (without passing through it), up to
$t=\infty$.

First of all we shall investigate how the holographic dark energy
scenario may work in the cases of the other contender models for
dark energy, namely, the K-essence [17] and the generalized
Chaplygin gas [18] frameworks. In the case that instead of
quintessence we use K-essence then for the phantom energy regime
the scale factor is given by [19]
\[a=a_0\left(t-t_*\right)^{-2\beta/[3(1-\beta)]}, \]
where $a_0$ is an arbitrary initial value for the scale factor,
$t_*$ is the time at the big rip, which is also arbitrary in this
model, and $0<\beta<1$. The proper size of the event horizon would
again take on the infinite value
\begin{equation}
R_h=\frac{3(1-\beta)\left(t-t_*\right)^{-2\beta/[3(1-\beta)]}}{3-\beta}
\left.\left(\left(t-t_*\right)^{2\beta/[3(1-\beta)]+
1}\right)\right|^{\infty}_{t-t_*} =\infty .
\end{equation}
It is therefore possible to find particular sets of K-essence
parameters which lead to a scenario where mathematically consistent
fundamental theories can be formulated, such as it happens in
phantom quintessence. Also similar to the phantom quintessence case,
the holographic K-essence screen must also be placed at the scale of
the big rip singularity, $L=R_* =a(t)\int_t^{t_*}dt'/a(t')$.

We consider next a generalized Chaplygin gas. In this case we have
for the scale factor of the universe the relation [20]
\begin{equation}
\dot{a}=Ca\left(A+\frac{B}{a^{3(1+\alpha)}}\right)^{1/[2(1+\alpha)]}
,
\end{equation}
in which $C=\sqrt{8\pi G/3}$ and $A$, $B$ and $\alpha$ are
constant whose values generally specify the regime we are working
on [20]. It can then be written that
\begin{equation}
R_h=a\int_t^{\infty}\frac{dt}{a}=a\int_a^{\infty}\frac{da}{a\dot{a}}
.
\end{equation}
From Eqs. (5) and (6) the size of the event horizon can be
generally expressed by means of the integral equation
\begin{equation}
R_h= a\int_a^{\infty}\frac{da}{Ca^2\left(A+
\frac{B}{a^{3(1+\alpha)}}\right)^{1/[2(1+\alpha)]}} .
\end{equation}
Exact integration of Eq. (7) cannot be performed for general values
of the Chaplygin parameter $\alpha$. However, if we choose for the
equation of state of the generalized Chaplygin gas the expression
$p=A\rho^{1/3}$ (i.e. taking for $\alpha$ the value -1/3), the above
expression can be integrated in closed form to give
\begin{equation}
R_h=a\left.\left(\frac{1}{C^2\left(A+
\frac{B}{a^2}\right)^{1/2}}\right)\right|_a^{\infty} =\frac{a}{C^2
B}\left(A^{-1/2}-\left(A+\frac{B}{a^2}\right)^{-1/2}\right) .
\end{equation}
Now, in the regime of the Chaplygin-gas model where the dominant
energy condition is preserved [20], $B>0$, it can be seen that the
size of the event horizon increases with $a$ but keeps always a
finite value. In the case where the dominant energy condition is
violated [20], $B<0$, on the phantom regime, one again obtains the
same result, as it can be shown by taking into account that
$a>\sqrt{|B|/A}$ when $\alpha>-1$. That behavior must be
associated with the feature that Chaplygin models do not lead to a
big rip type singularity in the future [20,21]. This result can be
generalized to any allowed value of parameter $\alpha>-1$. Thus,
just like it happens in quintessence models with $w>-1$ or when
$B>0$, the size of the event horizon entering the holographic
Chaplygin phantom cosmology is always finite at finite $t$ and
hence, whereas $R_h$ would always enter the definition of
holographic Chaplygin gas Friedmann equation, both for $B>0$ and
$B<0$, no consistent fundamental theory based on a S-matrix could
be constructed in the Chaplygin scenario for any $B$.

\section{Is the equation of state of a phantom universe quantized?}

If the regions before and after the big rip are mutually connected
by tunneling circumventing the singularity, then the evolution of
an ideal observer in the universe would proceed as follows. It
will first evolve in an accelerating universe and sees how this
reaches a maximum finite size, then the observer enters e.g. a
traversable wormhole by the mouth opening at the expanding region
at a time $t_{enter}<t_*$ and shortcuts the space-time, so
circumventing the big rip singularity, to exit out from the
wormhole through its mouth opening up into the contracting region
at a nonzero time after the big rip ($t_{exit}>t_*$). Thereafter,
the observer will see how the universe will steadily contract down
to zero as time tends to infinity. For such an observer the
evolution of the universe does not reach any singularity at
$t=t_*$ but smoothly goes from any initial time to a infinite time
after short cutting the big rip singularity by traversing a
wormhole. Mere inspection of the time dependence of the scale
factor when $w<-1$ indicates however that not any value of $w<-1$
leads to a real value of $a(t)$. We now consider in some more
detail how the solution to the cosmic-fundamental theory puzzle
can be worked out in the quintessential and K-essence phantom
models. In the former case, it can be straightforwardly realized
that no all values for parameter $w<-1$ can be continued into the
region after the big rip. Indeed, in general all those values of
$w$ which do not satisfy the relation
\begin{equation}
w= -\frac{1}{3}\left(1+\frac{2n+3}{n+1}\right), \;\; n=0, 1, 2, ...,
\infty
\end{equation}
lead to a negative or imaginary scale factor after the big rip.

Generally one would expect the equation-of-state parameter $w$ to
vary smoothly as the dark energy evolves. In the regime where dark
energy dominates, this is certainly true when $w>-1$, but it is by
no means a physical requirement in case that $w<-1$ as we are going
to show in what follows.  Condition (9) restricts the phantom models
which can allow for a cosmic evolution after the big rip, and
therefore the models which can allow for the presence of wormholes
and ringholes in the neighborhood of the big rip. In terms of the
holographic parameter $c$ that condition is given by
\begin{equation}
c=\frac{2(n+1)}{2n+3},\;\;\; n=0, 1, 2, ..., \infty
\end{equation}
We note that the allowed absolute values of $c$ and $w$ decrease
with $n$ and tend both to 1 (the cosmological constant case) as
$n\rightarrow\infty$.

Condition (9) comes from the requirement that the scale factor
$a(t)$ be real and positive also on the region $t>t_*$ when
$w<-1$. This does not represent a true {\it a priori} quantization
of the scalar field that makes up phantom energy, but merely a
condition that quantizes the equation of state assumed to
approximately govern the whole content of the universe in the
sense that only some of the possible values of the index of such a
equation of state are allowed when the phantom energy largely
dominates over all other energies. However, from the definitions
of the energy density, $\rho$, and pressure, $p$, in terms of a
scalar field $\phi(t)$,
\begin{equation}
\rho=\frac{1}{2}\dot{\phi}^2+V(\phi)
\end{equation}
\begin{equation}
p=\frac{1}{2}\dot{\phi}^2-V(\phi) ,
\end{equation}
the condition (9) leads also to a somehow "quantized" phantom field
theory given by
\begin{equation}
\phi=i\phi_0 \sqrt{\frac{\rho_0}{3(n+1)D}}\ln a(t)
\end{equation}
\begin{equation}
V(\phi)=\frac{6n+7}{6(n+1)}\rho_0a^{1/(n+1)}
=\frac{6n+7}{6(n+1)}\rho_0
e^{-i\sqrt{\frac{3D}{(n+1)\rho_0}}\phi/\phi_0} ,
\end{equation}
where $D=8\pi G\rho_0/3$ and $\rho_0$ and $\phi_0$ are integration
constants. This may be viewed as the pre-quantized theory (that
is, a theory quantized at a similar level to e.g. Bohr's hydrogen
theory) associated with a cosmological complementarity principle,
namely that all objects in the phantom universe have two aspects
of their existence, on the one hand as members of the cosmic
collective which can most generally be characvterized by the scale
factor $a$, and on the other hand as local individuals in the
universe and hence can be made to depend on the scalar field
$\phi$. Pairs of numbers are therefore necessary to define the
value of each dynamic physical variable describing the state of
the phantom universe. The algebra of complex numbers may take care
of this, whereby the real component is always associated with the
collective aspect ($a$) and the imaginary component is associated
with the individualized aspect ($\phi$). This would ultimately
provide with an explanation to the hitherto unexplained fact that
the kinetic field term $\dot{\phi}^2$ is definite negative for
phantom energy, and would restrict the occurrence of wormholes and
ringholes on only the neighborhood of the big rip singularity. In
fact, it is well known that the catastrophic creation of particles
that concentrate on the chronology horizon near their throat
renders these tunnels unstable [22], so that all those wormholes
and ringholes naturally created in a phantom universe would be
unstable except on regions near the big rip where the created
particles cease to have any individualized behavior to all adhere
to the collective ripping apart produced by the super-accelerated
expansion that breaks down all forces leading to any concentration
of such particles on the chronology horizon.

The allowed proper sizes of the future event horizon which satisfy
the above condition (9) are given by
\begin{equation}
R_h=-\frac{2(n+1)}{2n+3}T(t)^{-2(n+1)}\left.\left(T(t')^{2n+3}\right)\right|_{t}^{\infty}
.
\end{equation}
Relative then to an observer at a time $t<t_*$, since the power of
$T(t')$ is definite odd and $T(t')$ is definite negative after
$t_*$, $R_h=\frac{2(n+1)}{2n+3}T(t)+\infty$. As the observer
approaches the big rip from $t<t_*$, $R_h$ will remain being
$+\infty$ and $\dot{R}_h=-(2n+3)^{-1}$, and at $t=t_*$, where
$T(t)=0$, $R_h=+\infty$ as well. For $t>t_*$, we will have
$R_h=+\infty-\frac{2(n+1)}{2n+3}|T(t)|$ which keeps being
infinity, up to $t\rightarrow\infty$ at which asymptotic situation
$R_h$ shrinks to zero, just like it does in case that $c>1$. It
follows that if the condition (9) is satisfied then the proper
size of the future horizon does not shrink from its infinite
value, all the way, even at $t=t_*$, up to $t=\infty$, at which
point it vanishes. In the case of the holographic phantom model,
it can be seen that the evolution of the universe along time
$t<t_*$ will progress so that the size of $R_*$ shrinks down to
just the size of a wormhole or ringhole mouth (which then measures
the IR cutoff) that exceeds the UV cutoff which should necessarily
be placed at the scale of the hole throats, or less. This implies
that the IR cutoff will always remain larger than the UV cutoff at
any finite time in the future and therefore the definition of
holographic phantom energy is kept intact if condition (9) holds.
Hence, the global argument against holographic phantom raised by
Huang and Li [23] no longer applies if condition (9) is satisfied.
On the other hand, neither the argument by the same authors [23]
of a decreasing entropy could be adduced against holographic
phantom. In fact, if the entropy of the universe when only
holographic phantom energy is present is given by
\begin{equation}
S_{ph}=\pi M_p^2 R_* ,
\end{equation}
then the feature that $\dot{R}_h=c-1=-(2n+3)^{-1}<0$ implies that
$S_{ph}$ decreases as $t$ goes on. However, since the temperature
of the phantom stuff is definite negative [24], processes for
which the entropy decreases would be physically allowed. All the
above discussion can also be applied to the case of phantom
K-essence by taking into account that in this case the condition
that the parameter $\beta$ should satisfy is
\begin{equation}
\beta=\frac{3n}{3n+1}, \;\;\; n=1, 2, 3, ...
\end{equation}
with $n$ necessarily being finite as $\beta<1$. Using Eq. (17) we in
fact recover $R_h=+\infty$ and obtain
\[R_*=\frac{t_*-t}{2n+1} ,\]
for the K-essential phantom model.

\section{Circumventing the big rip with wormholes}

The solution to the cosmic puzzle for fundamental theories
provided by phantom energy has been already shown to require a
connection between the regions before and after the big rip which
can be thought to be implemented by means of Lorentzian wormholes
and ringholes [15]. Such a tunneling process would lead to a
multiply connected effective cosmic space-time ranging from $t=0$
to $t=\infty$, with the holes taking place everywhere on both
sides of the big rip singularity which is whereby avoided by any
signaling. Actually, wormholes and ringholes are quite natural
objects in the presence of phantom energy, a situation where the
dominant energy condition is violated. In that situation one would
expect the phantom stuff itself to make the exotic material
required for these tunnels to avoid pinching off and be really
traversable [16]. In the neighborhood of the big rip, i.e. on the
region where the tunnels are stable even quantum-mechanically, $T$
becomes so small that we can in fact take
\begin{equation}
a(t)=\frac{1}{T^{2(n+1)}}\simeq\frac{a_0}{\sin\left(T^{2(n+1)}/a_0\right)}=\frac{a_0}{\sin
x} ,
\end{equation}
so that in the considered region the metric becomes
\begin{equation}
ds^2=-dt^2 +\frac{a_0^2}{\sin^2 x}d\Omega_3^2 ,
\end{equation}
with $d\Omega_3^2$ the metric on the unit three-sphere. This
space-time can also be visualized as a five-hyperboloid defined by
\begin{equation}
-x_0^2+\sum_{j=1}^{4} x_j^2=a_0^2 .
\end{equation}
This hyperboloid can be embedded in $E^5$ so that the most general
approximate expression for the metric of the phantom accelerating
space in the neighborhood of the big rip is then that which is
induced in this embedding, that is,
\begin{equation}
ds^2=-dx_0^2 +\sum_{j=1}^{4}dx_j^2 ,
\end{equation}
whose topology is $R\times S^4$ and invariance group can be
approximated by the group $SO(4,1)$, showing ten Killing vectors
(four boosts and six rotations).

Metric (21) can be exhibited in a not still static form by
introducing the set of specific coordinates $x\in(0,\infty)$,
$\Psi_3,\Psi_2\in(0,\pi)$, $\Psi_1\in(0,2\pi)$,
\[x_4=a_0\sin\Psi_3\sin\Psi_2\cos\Psi_1 \]
\[x_3=a_0\sin\Psi_3\sin\Psi_2\sin\Psi_1 \]
\begin{equation}
x_2=a_0\sin\Psi_3\cos\Psi_2
\end{equation}
\[x_1=a_0\frac{\cos\Psi_3}{\sin x} \]
\[x_0=a_0\frac{\cos\Psi_3\cos x}{\sin x} \]

In terms of these coordinates, the above metric becomes
\begin{equation}
ds^2=-\left(1-a_0^{-2}r^2\right)\frac{a_0^2 C^2\rho_0 dt^2}{T(t)^2}
+\frac{dr^2}{\left(1-a_0^{-2}r^2\right)}+r^2 d\Omega_2^2 ,
\end{equation}
in which we have taken $r=a_0\sin\Psi_3$ and $d\Omega_2^2$ is the
metric on the unit two-sphere. This metric shows both the apparent
singularity at an event horizon at $r=a_0$ and the curvature
singularity at the big rip $t=t_*$. The latter singularity can
however be removed from the metric, so converting this into a
static metric, if we redefine time so that it will cover the
entire interval from $-\infty$ to $+\infty$. With the new time
coordinate
\begin{equation}
\tau=\ln\left[\left(\frac{T(t)}{T_0}\right)^{-2(n+1)}\right] ,
\end{equation}
where $T_0$ is an integration constant, we in fact get a static
metric having exactly the form of the De Sitter line element:
\begin{equation}
ds^2=-\left(1-a_0^{-2}r^2\right)d\tau^2
+\frac{dr^2}{\left(1-a_0^{-2}r^2\right)}+r^2 d\Omega_2^2 ,
\end{equation}
in the neighborhood of the big rip singularity. Now, in order to
exhibit that region as a multiply connected region due to the
presence of wormholes and ringholes, we again visualize the
space-time by the five-hyperboloid (20), with embedding metric (21),
exhibiting it by means of coordinates (22), but with $x_1$ and $x_0$
replaced for
\begin{equation}
x_1\rightarrow a_0\cos\Psi_3\cosh(\tau/a_0) ,\;\;\; x_0\rightarrow
a_0\cos\Psi_3\sinh(\tau/a_0) .
\end{equation}
One can then show that at least in the close neighborhood of the big
rip singularity the phantom space-time can be made to have a
multiply connected topology because a symmetry like that is
satisfied by the Minkowskian covering to Misner space is holding on
that region. In fact, on the Minkowskian five-hyperboloid
visualizing de Sitter space such a symmetry can be expressed by the
identifications
\begin{equation}
\left(x_0, x_1, x_2, x_3, x_4\right)\leftrightarrow
\left(x_0\cosh(mb)+x_1\sinh(mb),\right.
\left.x_1\cosh(mb)+x_0\sinh(mb), x_2, x_3, x_4\right) ,
\end{equation}
where $b$ is a dimensionless arbitrary constant and $m$ is any
integral number. There will be then a boost transformation in the
space described by metric (25) which is implied by the boost
transformation in the $(x_0, x_1)$-plane driven by the above
identification. It can be checked that metric (19) is invariant
under symmetry (27) in the region covered by that metric defined
by $x_1 > |x_0|$, with boundaries at $x_1 =\pm x_0$ (which
correspond to the big rip $t=t_*$ and $t=\infty$) and $x_4+x_3+x_2
=a_0^2$ (i.e. the event horizon at $r=a_0$ on the space-time with
metric characterized by time $\tau$) , provided
$\tau\rightarrow\tau+mb$. These boundaries describe the chronology
horizons for the region around the big rip. Now, since a
chronology horizon generally describes the onset of a non-chronal
region filled with closed timelike curves (CTC's), a region which
on the coordinates defined in terms of time $\tau$ goes from
$\tau=-\infty$ to $\tau=+\infty$, or vice versa, inside the event
horizon, would reflect into an also non-chronal region for
coordinates defined in terms of time $t$ whose onset is either at
$t=+\infty$ to go continuously down to $t=t_*$, after passing
through $t=0$, or at $t=t_*$ to continuously go up to $t=+\infty$,
after passing through $t=0$ as well. The existence of CTCs in the
non-chronal region can be implemented by the presence of naturally
existing grown up [21] wormholes and ringholes which then could,
according to the above discussion, mutually connect the
non-chronal regions before and after the big rip, such as it has
been depicted in Fig. 1.

\begin{figure}
\includegraphics[width=.9\columnwidth]{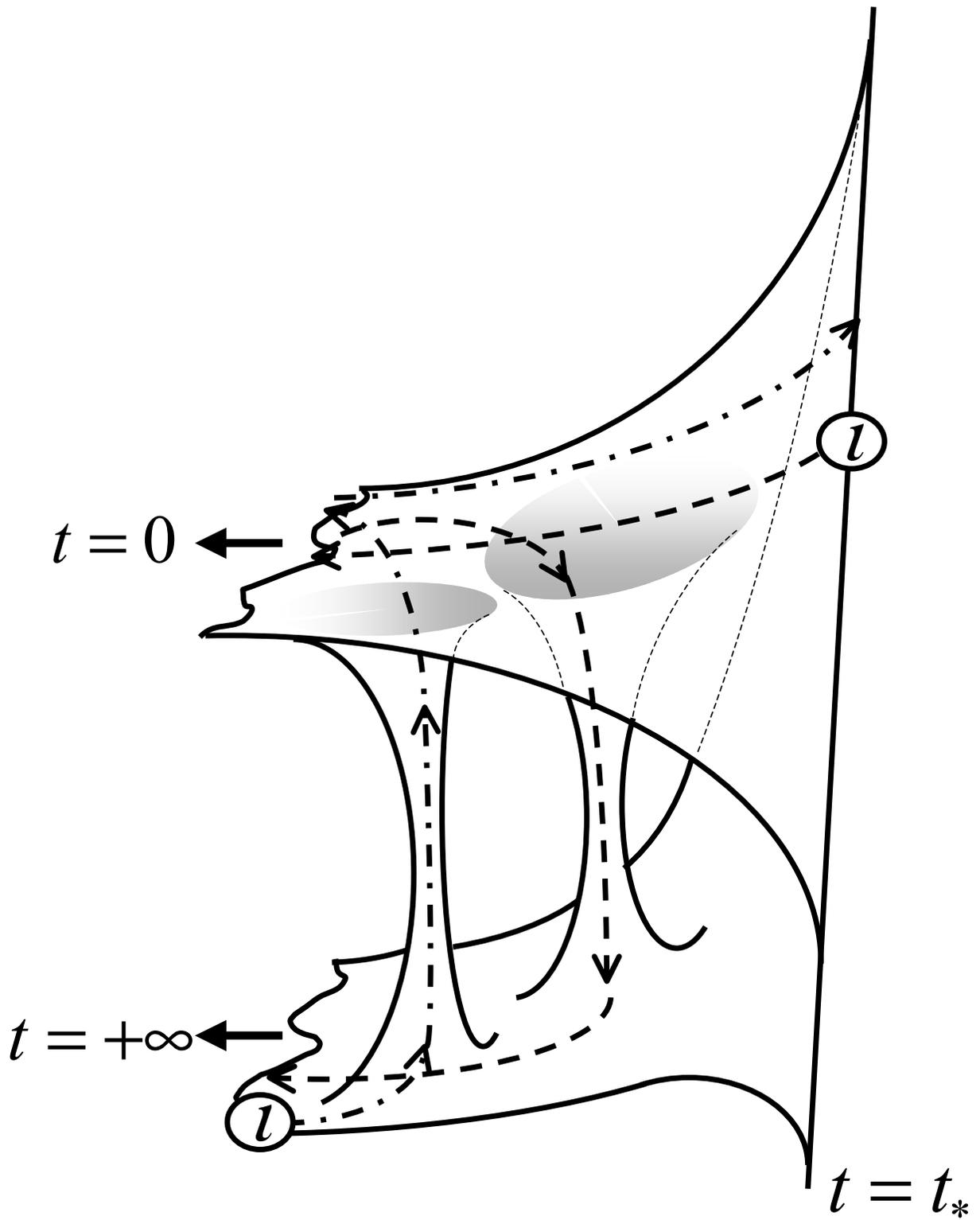}
\caption{\label{fig:epsart} Pictorial representation of the
possible trajectories that signalings may follow through wormholes
connecting the space-time regions before and after the big rip
singularity. The $\iota$s on the figure label the origin of the
signals.}
\end{figure}

For phantom cosmological models which do not satisfy condition (9)
[or (10)] the connection between both sides of the singularity could
not be made as the universe would cease to exist as a real entity,
just after the big rip. In such a case, the largest possible proper
size of the future horizon would be given by $R_*=cT(t)<R_h(c\geq
1)$ ,with $c\neq\frac{2(n+1)}{2n+3}$, which is necessarily finite,
so keeping or even aggravating the puzzle on fundamental theories.
We may therefore conclude that if we want to have an accelerating
cosmology which be free from the critical problems posed to the
definition of fundamental theories by the existence of a finite
horizon in the future, then we have to recourse to a phantom
quintessence or K-essence model restricted by the condition (9) or
(17). We notice that that condition in turn implied a minimal value
for $w$ of $w_{min}= -4/3$ and actually a quantization of the index
of the equation of state of the universe, according to which any
feasible tracking model of the universe with a future big rip
singularity would undergo an evolution that proceeded by steps and
be in this way governed by some still unexplored quantum rules.

Avoiding the big rip by tunneling through wormholes does not
actually contradict the idea that the holographic surface be
placed at the big rip as it could at first sight seem. In fact,
according to the covariant definition of a holographic surface
[25], the horizon at the big rip would correspond to an optimal
holographic screen both in the absence of closed timelike curves
near the singularity and in the presence of wormholes around that
singularity, as in both cases this surface marks the place where
the generating geodesics cease to show expansion. If there are not
closed timelke curves the holographic big rip surface would
represent the end of the universe, but there will be a contracting
region after that surface if closed timelike curves are allowed to
occur near it. This can be better seen on the corresponding
Penrose-Carter diagrams (see Fig. 2). On these diagrams, it can be
shown that, according to the Bousso's covariant holographic
formulation [25], the holographic optimal screen is always placed
on the big rip, no matter whether or not there are wormholes on
both sides of the singularity.

\begin{figure}
\includegraphics[width=.9\columnwidth]{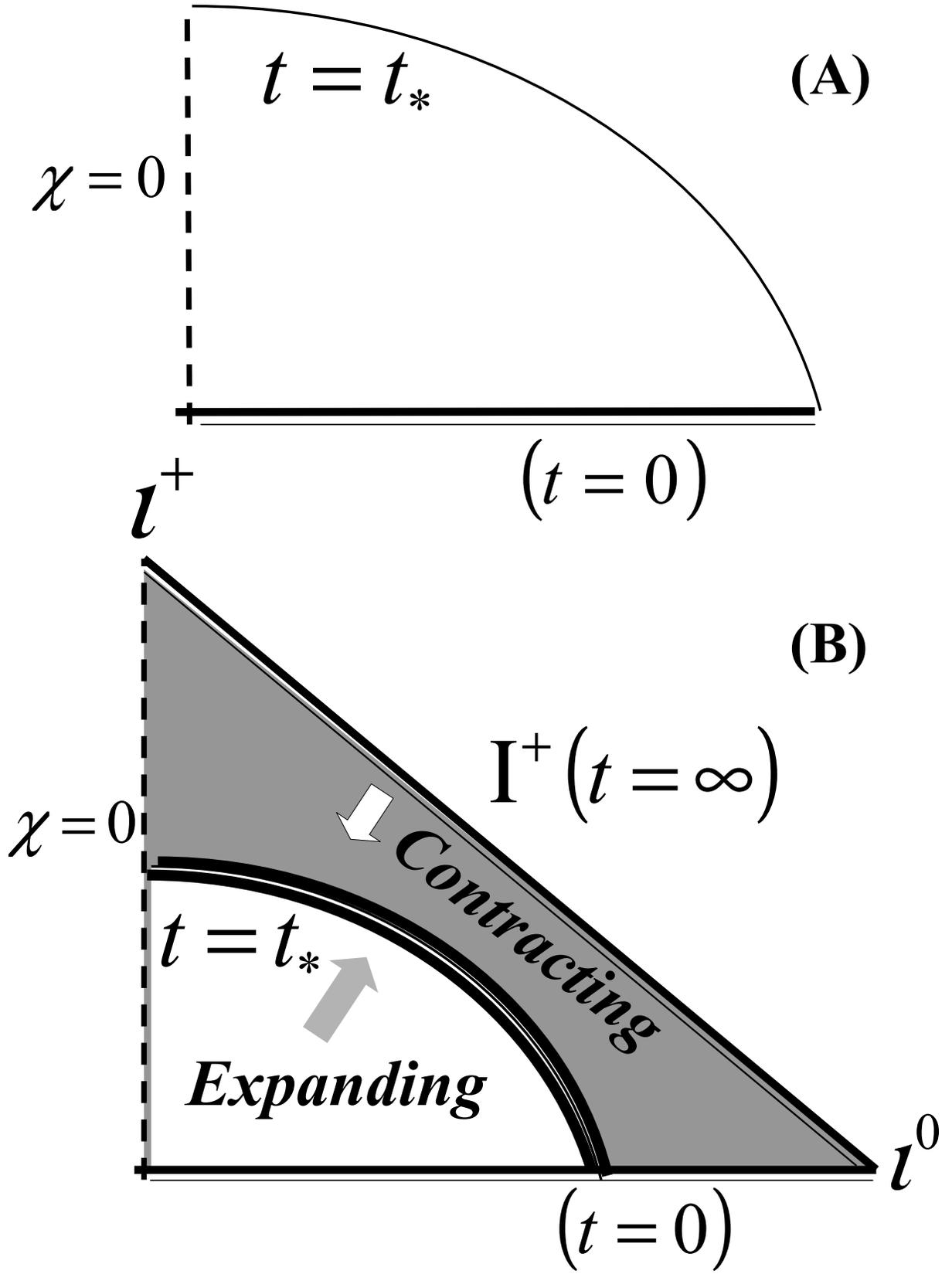}
\caption{\label{fig:epsart} Penrose-Carter diagrams and their
corresponding holographic screen for a phantom universe with (A)
no wormholes and (B) wormholes which continuously branch off from
and to the space-time regions very near to the big rip
singularity. The arrows indicate the direction of geodesics with
positive expansion. In case (B) the screen surface becomes a
screen volume (see the text). }
\end{figure}

The above possibility to avoid the big rip actually involves all
observers in the universe because of the following two reasons.
(1) Phantom energy is expected to largely dominate on the
neighborhood of the big rip singularity, and therefore wormholes
having as their exotic stuff energy that phantom energy should
crop up everywhere on that neighborhood; and (2) the Bekenstein
information-energy bound [26],
\[I<I_{max}=\frac{2\pi ER}{\hbar c \log 2} ,\]
from which the holographic principle truly originates, would
dictate that for a phantom universe with wormholes around the big
rip the spherical radius [12] $R=a(t)\int_t^{t_{max}}
dt'/a(t')=\infty$, and hence $I_{max}=\infty$, that is there
exists no restriction on the amount of information that can be
transferred through the wormholes from one side of the big rip
singularity to the other.

It is worth noticing that the existence of CTCs in the
neighborhood of the big trip hypersurface on both sides of
$t=t_*$, may extend and modify the Bousso's definition of that
hypersurface as a holographic surface [25]. Actually, screens,
preferred screens and optimal screens are geometrical concepts
which can only be defined in a causal space-time. If we would not
allow CTCs to occur then the surface at the big rip, let's denote
it by $B$, corresponded to a preferred screen in the Bousso's
sense, as the geodesic generators of it in the accelerated
expansion ceases to have positive expansion and turns to be
negative precisely at the big rip. Nevertheless, when CTCs are
allowed to occur there will also be null geodesics which turn from
positive to negative expansion before or after reaching surface
$B$, albeit very near to it. This would ultimately leads to the
notion of preferred or optimal screens having a given width, and
hence of a D-dimensional holographic bound where the information
is encoded not just on a D-1 dimensional surface but on a D
dimensional volume with small width.

\section{Further comments}

The question now is, does this quantization of the equation of
state imply any quantization for the phantom energy?. Even though
the absence of asymptotic flatness in the cosmological phantom
space precludes a meaningful definition of energy for that space,
one may still recourse to its thermodynamic properties to obtain
an estimate of the phantom energy. In fact, from the general
definition of temperature and entropy for a dark energy fluid
[24], $T=\kappa (1+w)a^{-3w}$, $S=C_0\left(T/(1+w)\right)^{1/w}V$,
where $\kappa$ and $C_0$ are positive constants and $V$ is the
volume being considered, we can estimate
\begin{equation}
E\simeq ST =\kappa^{(1+w)/w}C_0 (1+w)a^{-3(1+w)}V .
\end{equation}
For the phantom regime, $w<-1$, we then have a negative quantized
energy given by
\begin{equation}
E_n =-\frac{C(n)}{3(n+1)}a^{1/(n+1)}V,
\end{equation}
where the constant $C(n)$ is given by
\begin{equation}
C(n)=\gamma C_0\kappa^{3/(3n+4)} ,
\end{equation}
with $\gamma$ a numerical constant. For a box with given volume
$V$ filled only with phantom energy at a given time $t$, the
negative phantom energy is bounded from above for $n=\infty$, at
$E_{\infty}=0$. Moreover, since $T$ and $E_n$ are both definite
negative, the population probability law will have the usual
Boltzmann dependence, $\exp(-|E(n)|/k_b |T|)$, and there will be
no population inversion. If we immerse a given matter level-system
$\sigma$ into the box filled with phantom energy, since the energy
of the system is positive, then any energy exchange between the
system $\sigma$ and the phantom fluid would inexorably lead to a
decrease of the energy of $\sigma$, and hence the energy of that
system will be also bounded from above. We see therefore that the
conditions for the existence of a negative temperature (that is
the quantum nature of the whole system and energy boundedness from
above) are satisfied, contrary to the claim in Ref. 27.

Even more interesting is the fact that at a given $t$ any amount of
phantom energy enclosed in a volume $V$ can only change by jumps,
the energy exchanged in each jump $n\rightarrow n-1$ being given by
\begin{equation}
\Delta E_{ph}=\frac{(n+1)C(n-1)a^{1/n}-nC(n)a^{1/(n+1)}}{3n(n+1)} V
.
\end{equation}
Thus, processes such as the decay of phantom fields into gravitons
which were considered to be responsible for the violent
instabilities of the phantom energy [28] can no longer take place,
so that phantom energy is rendered stable. In this way, although
the phantom shows rather weird properties, none of them seems to
be precluding its existence in a cosmological context where most
observations really point to values of $w$ less than -1. The
almost desperate attempts to relax the observed values of
parameter $w$ from $w<-1$ to $w>-1$ by invoking mimicking illusion
effects [29,30] or the conversion of photons emitted by SNe Ia
into ultra-light axions [31] have proved not to be enough in
magnitude or generality (by defect or excess) to justify their
purposes. After all, if phantom energy is quantized one could
expect most, if not all, of its "bad" properties to be justified.

\acknowledgments

\noindent The author thanks Carmen L. Sig\"{u}enza and Alberto Rozas
for useful conversations. This work was supported by DGICYT under
Research Project BMF2002-03758.

\end{document}